\documentclass[floatfix,twocolumn,aps,showpacs,showkeys,nofootinbib]{revtex4}

\usepackage{amssymb}
\usepackage{amsmath}
\usepackage{graphics}
\usepackage{epsfig}
\usepackage[dvips]{color}
\usepackage{bm}
\usepackage{color}

\newcommand{\be}{\begin{equation}}

\newcommand{\ee}{\end{equation}}
\newcommand{\bea}{\begin{eqnarray}}
\newcommand{\eea}{\end{eqnarray}}
\newcommand{\beas}{\begin{eqnarray*}}
\newcommand{\eeas}{\end{eqnarray*}}

\newcommand{\nn}{\nonumber\\}

\begin{document}
\title{Inverse magnetic catalysis for the chiral transition induced by thermo-magnetic effects on the coupling constant}
\author{Alejandro Ayala$^{1,3}$, M. Loewe$^{2,3}$, Ana J\'ulia Mizher$^1$, R. Zamora$^2$}
\affiliation{$^1$Instituto de Ciencias
  Nucleares, Universidad Nacional Aut\'onoma de M\'exico, Apartado
  Postal 70-543, M\'exico Distrito Federal 04510,
  Mexico.\\
  $^2$Instituto de F\1sica, Pontificia Universidad Cat\'olica de Chile,
  Casilla 306, Santiago 22, Chile.\\
  $^3$Centre for Theoretical and Mathematical Physics, and Department of Physics,
  University of Cape Town, Rondebosch 7700, South Africa}

\begin{abstract}

We compute the one-loop thermo-magnetic corrections to the self-coupling in a model where charged scalars interact also with a constant magnetic field. The calculation is motivated by the possibility that the critical temperature for the chiral phase transition in a magnetic background can be influenced by the dependence of the coupling constant on the magnetic field. We show that the coupling decreases as a function of the field strength. This functional dependence introduces in turn a correction to the boson masses which causes the critical temperature to decrease as a function of the field strength. 

\end{abstract}



\maketitle

In recent years there has been a great  interest on the effects that a magnetic field may have on the QCD phase transitions. Peripheral heavy-ion collisions offer the possibility to test these effects since intense fields are generated, particularly during the very early stages of these collisions~\cite{Kharzeev, Skokov, Bzdak}. 

In the past lattice simulations, performed with unphysical large pion masses, indicated that the critical temperature increased with the intensity of the field ~\cite{Braguta, D'Elia:2011zu, D'Elia2}. Nowadays, the consensus is that the critical temperature shows the opposite behavior, namely, it decreases with the intensity of the field~\cite{Fodor,Bali:2012zg}. The first model results also showed an increasing critical temperature with increasing field~\cite{Loewe:2013coa, Agasian:2008tb, Fraga:2008qn, Mizher:2010zb}.

Several ideas have been put forward to explain this surprising behavior, dubbed {\it inverse magnetic catalysis} (though a more appropriate name could be magnetic anticatalysis, given that the field acts against the formation of the condensate): Ref.~\cite{Noronha} argues that if the fermion contribution to the pressure is paramagnetic, with a sufficiently large magnetization, the critical temperature decreases with the magnetic field strength. Refs.~\cite{Bruckmann:2013oba} and~\cite{Ferreira} attribute such a decrease to a back reaction of the Polyakov loop, which indirectly feels the magnetic field. More recently, Refs.~\cite{Farias, Ferreira1} have postulated an ad hoc magnetic field and temperature dependent running coupling, inspired by the QCD running of the coupling with energy, in the Nambu-Jona-Lasinio model, which makes the critical temperature decrease with increasing magnetic field. 

Given the variety of explanations for the inverse magnetic catalysis one can ask for the veracity of the commonly held belief that both aspects of QCD, namely deconfinement and chiral symmetry breaking, need to be present together in order to understand the phenomenon. Here we explore one of these aspects, namely chiral symmetry, within a scalar model, the Abelian Higgs model. We show that inverse magnetic catalysis happens in this case provided the coupling constant is consistently corrected within the model to include thermo-magnetic effects. The essential ingredients are the finite temperature effective potential in the presence of a magnetic field and the proper handling of plasma screening effects that have been recently consistently formulated within the model at hand~\cite{ahmrv}. Indeed, in a recent work we have explored the behavior of the critical temperature for the chiral transition in an effective model with spontaneous symmetry breaking where charged bosons and fermions interact with a constant magnetic field~\cite{ahmrv}. The calculation accounted for the magnetic corrections introduced to the effective potential but otherwise considered constant couplings. In this work we use the Abelian Higgs model to explore the scenario where the self-coupling receives thermo-magnetic modifications to one-loop order and its effects on the critical temperature for the phase transition. 

The model is given by the Lagrangian 
\bea
   {\mathcal{L}}=(D_{\mu}\phi)^{\dag}D^{\mu}\phi
   +\mu^{2}\phi^{\dag}\phi-\frac{\lambda}{4}   
   (\phi^{\dag}\phi)^{2},
\label{lagrangian}
\eea
where $\phi$ is a charged scalar field and
 \bea
   D_{\mu}=\partial_{\mu}+iqA_{\mu},
\label{dcovariant}
\eea
is the covariant derivative. $A^\mu$ is the vector potential corresponding to an external magnetic field directed along the $\hat{z}$ axis,
\bea
   A^\mu=\frac{B}{2}(0,-y,x,0),
\label{vecpot}
\eea
and $q$ is the particle's electric charge. $A^\mu$ satisfies the gauge condition $\partial_\mu A^\mu=0$. In the language of the covariant $R_\xi$ gauges, this gauge condition corresponds to $\xi=0$ and therefore the Goldstone mode does couple to the gauge field. Since the gauge field is taken as classical, we do not consider its fluctuations and thus no loops involving the gauge field in internal lines. The squared mass parameter $\mu^2$ and the self-coupling $\lambda$ are taken to be positive.

We can write the complex field $\phi$ in terms of the components $\sigma$ and $\chi$,
\bea
   \phi(x)&=&\frac{1}{\sqrt{2}}[\sigma(x)+i \chi(x)],\nn
   \phi^{\dag}(x)&=&\frac{1}{\sqrt{2}}[\sigma(x)-i\chi(x)].
\label{complexfield}
\eea
To allow for an spontaneous breaking of symmetry, we let the $\sigma$ field to develop a vacuum expectation value $v$
\bea
   \sigma \rightarrow \sigma + v,
\label{shift}
\eea
which can later be taken as the order parameter of the theory. After this shift, the Lagrangian can be rewritten as
\bea
   {\mathcal{L}} &=& -\frac{1}{2}[\sigma(\partial_{\mu}+iqA_{\mu})^{2}\sigma]-\frac{1}
   {2}\left(\frac{3\lambda v^{2}}{4}-\mu^{2} \right)\sigma^{2}\nn
   &-&\frac{1}{2}[\chi(\partial_{\mu}+iqA_{\mu})^{2}\chi]-\frac{1}{2}\left(\frac{\lambda v^{2}}{4}-   
   \mu^{2} \right)\chi^{2}+\frac{\mu^{2}}{2}v^{2}\nn
  &-&\frac{\lambda}{16}v^{4}
  +{\mathcal{L}}_{I},
  \label{lagranreal}
\eea
where ${\mathcal{L}}_{I}$ is given by
\bea
  {\mathcal{L}}_{I}&=&-\frac{\lambda}{16}\left(\sigma^4+\chi^4+2\sigma^2\chi^2\right),
  \label{lagranint}
\eea
and represents the Lagrangian describing the interactions among the fields $\sigma$ and $\chi$, after symmetry breaking. It is well known that in the Abelian Higgs model the gauge field $A^\mu$ acquires a finite mass and thus cannot represent the physical situation of a massless photon interacting with the charged scalar field. Therefore, for the discussion we ignore the mass generated for $A^\mu$ as well as issues regarding renormalization after symmetry breaking and concentrate on the scalar sector. From Eq.~(\ref{lagranreal}) we see that the $\sigma$ and $\chi$ masses are given by
\bea
  m^{2}_{\sigma}&=&\frac{3}{4}\lambda v^{2}-\mu^{2},\nn
  m^{2}_{\chi}&=&\frac{1}{4}\lambda v^{2}-\mu^{2}.
\label{masses}
\eea
\begin{figure}[t!]
\begin{center}
\includegraphics[scale=0.5]{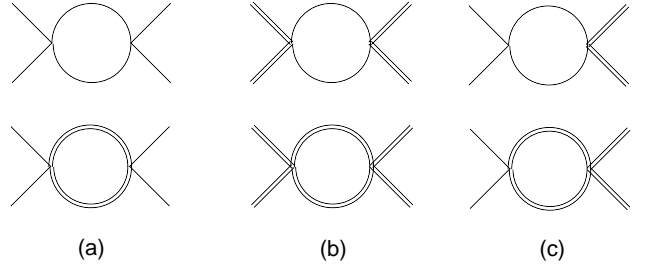}
\end{center}
\caption{One-loop Feynman diagrams that contribute to the thermal and magnetic correction to the coupling $\lambda$. Single (double) lines represent the $\sigma$ ($\chi$) field.}
\label{fig1}
\end{figure}
Including the $v$-independent terms and choosing the renormalization scale as $\tilde{\mu}=e^{-1/2}\mu$, it has been shown in Ref.~\cite{ahmrv} that the thermo-magnetic effective potential in the small to intermediate field regime can be written as
\bea
   V^{({\mbox{\small{eff}}})}&=&
   -\frac{\mu^2}{2}v^2 + \frac{\lambda}{16}v^4\nn
   &+&\sum_{i=\sigma,\chi}\left\{\frac{m_i^4}{64\pi^2}\left[ \ln\left(\frac{(4\pi T)^2}{2\mu^2}\right) 
   -2\gamma_E +1\right]
   \right.\nn
   &-&\frac{\pi^2T^4}{90} + \frac{m_i^2T^2}{24}\nn
   &+&\frac{T(2qB)^{3/2}}{8\pi}\zeta\left(-\frac{1}{2},\frac{1}{2}+\frac{m_i^2+\Pi}{2qB}\right)\nn
   &-&\frac{(qB)^2}{192\pi^2}\left[ \ln\left(\frac{(4\pi T)^2}{2\mu^2}\right)
   - 2\gamma_E + 1\right.\nn
   &+&\left.\left. 
   \zeta (3)\left(\frac{m_b}{2\pi T}\right)^2 - \frac{3}{4}\zeta (5) \left(\frac{m_b}{2\pi T}\right)^4   
   \right]\right\},
   \label{Veff-mid}
\eea
where $\gamma_E$ is Euler's gamma and we have introduced the leading temperature plasma screening effects for the boson's mass squared, encoded in the boson's self-energy
\bea
   \Pi=\lambda T^2/12.
\label{self}
\eea
For the Hurwitz zeta function $\zeta(-1/2,z)$ in Eq.~(\ref{Veff-mid}) to be real, we need that 
\bea
   -\mu^2 + \Pi > qB,
\label{requirealso}
\eea
condition that comes from requiring that the second argument of the Hurwitz zeta function satisfies $z>0$, even for the lowest value of $m_b^2$ which is obtained for $v=0$. Furthermore, for the large $T$ expansion to be valid, we also require that
\bea
   qB/T^2 <1.
\label{otherrequirement}
\eea

Let us now compute the one-loop correction to the coupling $\lambda$, including thermal and magnetic effects. Figure~\ref{fig1} shows the Feynman diagrams that contribute to this correction. Columns (a), (b) and (c) contribute to the correction to the $\sigma^4$, $\chi^4$ and $\sigma^2\chi^2$ terms of the interaction Lagrangian in Eq.~(\ref{lagranint}), respectively. Since these corrections are equivalent, we concentrate on the diagrams in column (a). Each of the two diagrams involves two propagators of the same charged boson, therefore they can be obtained from the general expression
\bea
   I(P_i;m_i^2)=\lambda^2T\sum_n\int\frac{d^3k}{(2\pi)^3}D_B(P_i-K)D_B(K),
\label{general}
\eea
where $P_i$ is the total incoming four-momentum and $D_B$ is the charged boson propagator in the presence of the magnetic field. Hereafter, capital letters are used to denote four-momenta in Eucledian space, {e.g.} $K\equiv (\omega_n,{\mbox{\bf{k}}})$, where $\omega_n=2n\pi T$ are boson Matsubara frequencies. The propagator $D_B$ is written using Schwinger's proper-time method and is given by
\bea
   D_B(K;m_i^2)=\int_0^\infty ds \frac{e^{-s(\omega_n^2+k_3^2+k_\perp^2\frac{\tanh (qBs)}{qBs} + m_i^2)}}{\cosh (qBs)}.
\label{Schwinger}
\eea
\begin{figure}[t!]
\begin{center}
\includegraphics[scale=0.5]{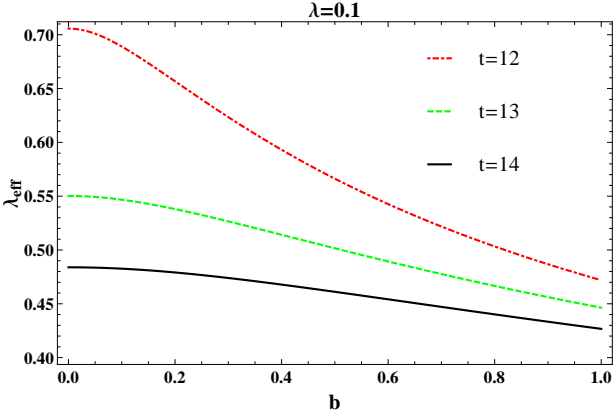}
\end{center}
\caption{Color on-line. $\lambda_{\mbox{\small{eff}}}(v=0)$ as a function of $b=qB/\mu^2$ for three different values of $t=T/\mu$.}
\label{fig2}
\end{figure}
Let us work in the {\it infrared limit}, namely, $P_i=(0,{\mbox{\bf{p}}}\rightarrow 0)$. It has been shown~\cite{reloaded} that this limit can be formally implemented by straightforward setting $P=0$ in Eq.~(\ref{general}). We work with the hierarchy of scales where $qB,m_i^2< T^2$. It is known that in order to properly implement this hierarchy~\cite{alhmrv}, it is necessary to separate the contribution from the zero mode from the rest of the modes in Eq.~(\ref{general}). We therefore write
\bea
   I(0;m_i^2)&\equiv& I_{n=0}(0;m_i^2) + I_{n\neq 0}(0;m_i^2)\nn
   I_{n\neq 0}(0;m_i^2)&=&\lambda^2T\sum_{n\neq 0}\int\frac{d^3k}{(2\pi)^3}
   D_B(\omega_n,{\mbox{\bf{k}}})D_B(\omega_n,{\mbox{\bf{k}}})\nn
   I_{n=0}(0;m_i^2)&=&\lambda^2T\int\frac{d^3k}{(2\pi)^3}D_B({\mbox{\bf{k}}})D_B({\mbox{\bf{k}}})
\label{separate}
\eea
$I_{n=0}$ is straightforward computed with the result
\bea
   I_{n=0}(0;m_i^2)=\frac{T}{16\pi}\frac{1}{(2qB)^{1/2}}  \zeta \left(\frac{3}{2}, \frac{1}{2} + \frac{m^2_{i}+\Pi}{2qB}\right),
\label{n=0}
\eea 
where we have also included the plasma screening effects for the boson's mass squared~\cite{ahmrv}. The contribution from the rest of the modes is performed by resorting to the weak field limit of the boson propagator~\cite{mexicanos}
\bea
   D_B(\omega_{n\neq 0},{\mbox{\bf{k}}})&=&\frac{1}{\omega_n^2+{\mbox{\bf{k}}}^2+m_i^2}
   \left[1-\frac{(qB)^2}{(\omega_n^2+{\mbox{\bf{k}}}^2+m_i^2)^2}\right.\nn
   &+&\left.\frac{2(qB)^2\ k_\perp^2}{(\omega_n^2+{\mbox{\bf{k}}}^2+m_i^2)^3}
   \right].
\label{propnneq0}
\eea
The sum and integrals in $I_{n\neq 0}$ are performed by means of the Mellin summation technique~\cite{Bedingham} with the result
\bea
   I_{n\neq 0}(0;m_i^2)&=&-\frac{1}{16 \pi^2} \Big[\ln\left( \frac{(4\pi T)^2}{2\mu^2} \right) +1 - 2\gamma_E\nn
   &+&\zeta(3) \left(\frac{\sqrt{m_i^2+\Pi}}{2 \pi T}\right)^2   \Big]\nn
   &-&\frac{(q B)^2 }{512 \pi^6 T^4} \zeta(5) \left( 1 -  \frac{1}{32 \pi} \right).
\label{nneq0}
\eea
\begin{figure}[t]
\vspace{0.4cm}
\begin{center}
\includegraphics[scale=0.34]{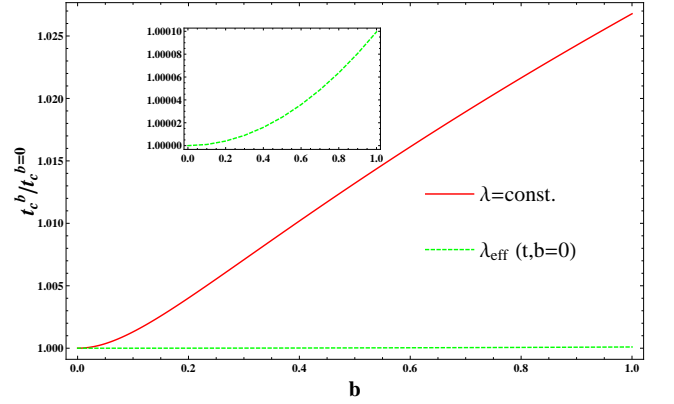}
\end{center}
\caption{Color on-line. Effect of the self-coupling on the critical temperature.  The solid curve corresponds to the case where the coupling is a constant $\lambda=0.9$. The dashed curve corresponds to the calculation where $\lambda_{\mbox{\small{eff}}}(T,qB=0;\lambda =0.9)$. This last curve is also a growing function of $b=qB/\mu^2$, as can better be seen in the inset, though the growth is less strong than the case computed with a constant $\lambda$.}
\label{fig3}
\end{figure}
We have also included the plasma screening effects for the boson's mass squared and have carried out the mass renormalization introducing a counter term $\delta m^2=-1/\epsilon+\gamma_E-\ln(2\pi)$. The function $I(0;m_i^2)$ is therefore explicitly obtained by adding up Eqs.~(\ref{n=0}) and~(\ref{nneq0}).

Considering the permutation factors and the contribution from the $s$, $t$ and $u$-channels, the correction for the self-coupling $\lambda$ to one-loop order is given by
\bea
   \lambda_{\mbox{\small{eff}}}=\lambda
   \left[1 + 6\lambda\left( 9I(0;m_\sigma^2) + I(0;m_\chi^2) \right)
   \right].
\label{lambdaeff}
\eea
Note that $\lambda_{\mbox{\small{eff}}}$ depends on $v$ through the dependence on $m_i^2$. Let us furthermore take the approximation where we evaluate $\lambda_{\mbox{\small{eff}}}$ at $v=0$. The rationale is that we are pursuing the effect on the critical temperature which is the temperature where the curvature of the effective potential at $v=0$ vanishes. Figure~\ref{fig2} shows the behavior of $\lambda_{\mbox{\small{eff}}}(v=0)$ as a function of $b=qB/\mu^2$ for three different values of $t=T/\mu$.  Note that in all cases the effective coupling is a decreasing function of the magnetic field strength.

Let us now study the effect that the self-coupling has on the critical temperature. We first look at the cases where we set the coupling to its tree level value and where only thermal effects are included. Figure~\ref{fig3} shows the critical temperature in these cases, obtained from setting the second derivative of Eq.~(\ref{Veff-mid}) equal to zero at $v=0$, normalized to the critical temperature for vanishing magnetic field. Note that in both cases the critical temperature is an increasing function of the field strength though, when the thermal effects on the coupling are included, this increase is tamed. 

\begin{figure}[t!]
\begin{center}
\includegraphics[scale=0.397]{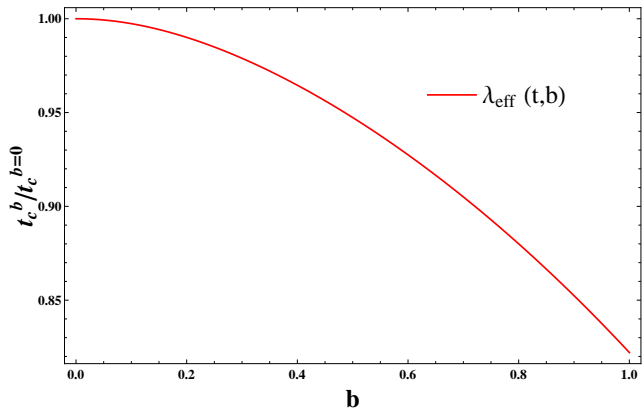}
\end{center}
\caption{Color on-line. Critical temperature computed with the full thermo-magnetic dependence of the self-coupling $\lambda_{\mbox{\small{eff}}}(T,qB;\lambda =0.9)$. In this case the critical temperature becomes a decreasing function of the field strength.}
\label{fig4}
\end{figure}

Figure~\ref{fig4} shows the critical temperature for the case where we consider the full thermo-magnetic dependence of the self-coupling. Note that in this case the critical temperature becomes a decreasing function of the field strength. 

In conclusion, we have shown that when including the one-loop thermo-magnetic effects for the self-coupling in a model where charged scalars interact with an external magnetic field and where the effective potential is computed beyond a mean field approximation considering the plasma screening properties, the critical temperature for the chiral transition becomes a decreasing function of the field strength. This behavior is a direct consequence of a decrease of the self-coupling with the field strength. There are two main differences between our approach and the approach in some earlier works~\cite{Andersen, Fukushima, Skokov2} that also go beyond a mean field approximation. On the one hand we have included in the analysis the plasma screening effects by means of the resummation of the so-called ring diagrams instead of resorting to a renormalization group analysis. This is particularly important for theories exhibiting spontaneous symmetry breaking where the boson masses can vanish and even become negative as functions of the order parameter. Second, the coupling that gets modified by magnetic field effects in the present work is the boson self-coupling whereas for instance in Ref.~\cite{Fukushima} it is a contact four-fermion interaction. We emphasize that the thermo-magnetic dependence of the self-coupling has been computed --as opposed to assumed-- within our working model. Though this model is not equivalent to QCD, it possesses one of its main features, namely, symmetry restoration at finite temperature whose primary physical consequence is the change of the particle's masses as a function of the order parameter. As follows from Eqs.~(\ref{masses}), since the self-coupling enters as an ingredient into the boson's masses, the modification of the self-coupling introduces a magnetic field dependence on these masses and thus on one of the main parameters describing symmetry restoration. It then seems that confinement may not necessarily be an ingredient needed to understand the lattice results, or at least that the magnetic field effects on the chiral symmetry restoration and confinement transitions may be independent. These findings will be further elaborated on by including the effects of fermions in an upcoming analysis.

\section*{Acknowledgments}

Support for this work has been received in part from DGAPA-UNAM under grant number PAPIIT-IN103811, CONACyT-M\'exico under grant number 128534 and FONDECYT under grant  numbers 1130056 and 1120770. R. Z. acknowledges support from CONICYT under Grant No. 21110295.

\end{document}